\documentstyle[12pt,subfigure]{article}
\input{psfig}

\newcommand   {\beq}       {\begin{equation}}
\newcommand   {\eeq}     {\end{equation}}
\newcommand   {\bea}     {\begin{eqnarray}}
\newcommand   {\eea}     {\end{eqnarray}}
\newcommand   {\rgyr}    {r_{\mbox{\tiny gyr}}}
\newcommand   {\ree}     {r_{\mbox{\tiny ee}}}
\newcommand   {\lika}    {\,=\,}
\newcommand   {\ev}[1]   {\langle #1\rangle}

\newcommand   {\Hmc}     {H_{\mbox{\tiny MC}}}
\newcommand   {\etal}    {{\it et~al.}}
\renewcommand {\ni}      {\noindent}

\newcommand{\AmS}{{\protect\the\textfont2
  A\kern-.1667em\lower.5ex\hbox{M}\kern-.125emS}}
\begin{document}

\vspace{48pt}
\begin{center}

{\Large\bf Studies of an off-lattice model for protein folding:\break\break
           Sequence dependence and improved sampling at\break\break    
           finite temperature}

\vspace{48pt}

{\sl Anders Irb\"ack}\footnote{irback@thep.lu.se.} 
and {\sl Frank Potthast}\footnote{frank@thep.lu.se.}

\vspace{12pt}

Department of Theoretical Physics, University of Lund\\
S\"olvegatan 14A, S-223 62 Lund, Sweden\\

\vspace{120pt}

{\large\bf Abstract}
\end{center}
\vspace{12pt}\ni

We study the thermodynamic behavior of a simple off-lattice
model for protein folding. The model is two-dimensional and
has two different ``amino acids''. Using numerical 
simulations of all chains containing eight or ten
monomers, we examine the sequence dependence at a fixed
temperature. It is shown that only a few of the chains exist in
unique folded state at this temperature, and the energy level spectra 
of chains with different types of behavior are compared. 
Furthermore, we use this model as a testbed for two improved Monte
Carlo algorithms. Both algorithms are based on
letting some parameter of the model become
a dynamical variable; one of the algorithms 
uses a fluctuating temperature and the other
a fluctuating monomer sequence. We find that by these algorithms 
one gains large factors in efficiency in comparison with
conventional methods.\\
\\
\\ 
\ni
Published in Journal of Chemical Physics 103, 10298--10305 (1995).
   
\vfill
\newpage
\section{INTRODUCTION}

The protein molecule is a very complex system, and understanding  
the folding of natural proteins remains one of the most challenging 
problems in biophysics. In recent years there has been an increasing 
interest in understanding the relevant mechanisms of the 
folding process by studying simplified models. Several authors have used 
ideas from spin-glass theory to gain useful insights into the
behavior of self-interacting random chains 
\cite{Stein:85,Bryngelson:87,Garel:88,Shakhnovich:89,Parisi:91}. 
The picture emerging from these studies is that the 
phase diagram consists of coil, globule and frozen phases.
In the frozen phase the system exists in one of a number of 
different states, each corresponding to a fairly well-defined 
shape. This phase is interesting from the viewpoint of proteins, but
for a generic random sequence it appears impossible to identify 
a state corresponding to the unique native state of a protein.   

The sequence dependence of the folding properties has been studied
numerically in various models
\cite{Go:78,Shakhnovich:91,Leopold:92,Miller:92,Fukugita:93,Sali:94,Shakhnovich:94,Chan:94,Bryngelson:95}. 
An extensive study was reported recently by \u{S}ali \etal~\cite{Sali:94},
who examined the behavior of 200 randomly selected sequences 
in a lattice model with contact interactions. It was
found that 30 of these exhibited a folded state which
was both thermodynamically dominant and kinetically 
accessible in a reasonable time. The existence of folding and nonfolding 
sequences has been observed also in off-lattice 
models~\cite{Fukugita:93}, but much less is known about 
the behavior of such models. In this paper we study the 
thermodynamic behavior of different sequences in a very simple 
off-lattice model.

The model studied has been proposed by Stillinger \etal~\cite{Stillinger:93}. 
It is two-dimensional and has only two kinds of ``amino acids''.  
In Refs.~\cite{Stillinger:93,Head-Gordon:93} the energy and structure of 
the ground state was determined for all possible chains with seven or 
fewer monomers, and an interesting interpretation of the results 
was obtained by employing neural-network techniques. In this 
paper we study the behavior of the model at finite temperature. 
To study the sequence dependence, we have carried out numerical simulations of 
all possible chains containing eight or ten monomers at a fixed
temperature. We find that only a few of the chains exist in 
a unique folded state at this temperature. These chains have
a relatively high folding temperature $T_f$. This property is important
because the dynamics tends to be very slow at low temperature.  
Chains with a high $T_f$ are therefore more likely to satisfy both 
the thermodynamic and kinetic requirements for folding. 
We have also determined the low-lying energy levels of 
a few different sequences. Sequences with good folding 
properties are expected to exhibit a large stability 
gap \cite{Bryngelson:95}, which is defined as the energy gap
between the native state and the lowest of all states with little
structural similarity to the native state. 
The stability gaps of the sequences we have studied show a wide
variation, and, as expected, it is large when $T_f$ is high.

This first part of our study shows that the model displays interesting
features for relatively short chains. It has been performed 
by using the well established hybrid Monte Carlo algorithm~\cite{Duane:87}. 
In the second part of the paper, we use the model as a testbed 
for two algorithms that are meant to facilitate the study of longer chains.
 
Simulations of heteropolymeric chains at low temperature
are notoriously difficult, due to the presence of high 
free-energy barriers separating different folded states. 
The problem is that any local algorithm 
requires the system to pass through these barriers, which leads to a 
suppression of transitions between different free-energy valleys. 
A method designed to overcome this difficulty is the multicanonical 
Monte Carlo algorithm~\cite{Berg:91}, which is closely 
related to the umbrella-sampling 
method~\cite{Torrie:77}. The trick used here is to simulate   
a modified energy function which one tries to choose so as to 
eliminate the free-energy barriers. The canonical distribution 
is then extracted by means of the reweighting 
technique~\cite{Ferrenberg:88}.

An alternative approach is provided by the method of 
simulated tempering \cite{Marinari:92}. In this method the temperature becomes 
a dynamical variable which takes values ranging over a definite set.   
In this way one tries to utilize the fact that at higher 
temperature the free-energy barriers are lower. In Ref.~\cite{Marinari:92} 
this method was successfully applied to the random-field Ising 
model. The idea to let some parameter of the model become 
a dynamical variable has subsequently been used to accelerate simulations
of other systems too~\cite{Kerler:93,Liu:93,Kerler:94a,Kerler:94b}.
It had earlier been shown by Lyubartsev \etal ~\cite{Lyubartsev:91} 
that this is a useful method for calculating the free energy.
 
In this paper we investigate the use of the dynamical-parameter 
method in simulating heteropolymeric chains. We consider the 
simulated-tempering algorithm, and also another algorithm of 
the same type which we call the multisequence algorithm. 
In this algorithm the sequence degrees 
of freedom become a dynamical variable, which means that a number of 
different monomer sequences are simulated in parallel. We  
compare the performance of these two algorithms with that of
hybrid Monte Carlo. Our results show that by the dynamical-parameter 
algorithms one greatly reduces the amount of computer time required 
for a representative sampling of the different folded states.

The plan of this paper is as follows. The model is described in Sec.~2
and the algorithms in Sec.~3. Sec.~4 deals with homopolymers, whereas        
general sequences are studied in Sec.~5. In Sec.~6 we present the results
of our tests of the dynamical-parameter algorithms. Sec.~7 is a summary. 

\section{THE MODEL}

The model studied has two kinds of monomers, to be called $A$ and $B$. 
The monomers are linked by rigid bonds of unit 
length to form linear chains living in two dimensions. 
For an $N$-mer we specify the sequence of monomers by the binary 
variables $\xi_1,\ldots,\xi_N$ and the configuration 
by the angles $\theta_2,\ldots,\theta_{N-1}$, where 
$\theta_i$ denotes the bend angle at site $i$ and is taken to
satisfy $|\theta_i|\le\pi$. The energy function is  
\beq
E(\theta,\xi)\lika \sum_{i=2}^{N-1} E_1(\theta_i) + 
\sum_{i=1}^{N-2}\sum_{j=i+2}^N E_2(r_{ij},\xi_i,\xi_j)
\label{e}\eeq   
where 
\bea
E_1(\theta_i)&=&{1\over 4}\,(1-\cos\theta_i) \nonumber \\
E_2(r_{ij},\xi_i,\xi_j)&=&
4(r_{ij}^{-12}-C(\xi_i,\xi_j)r_{ij}^{-6})
\label{e12}\eea
and $r_{ij}=r_{ij}(\theta_{i+1},\ldots,\theta_{j-1})$
denotes the distance between sites $i$ and $j$ of the chain. 
The term $E_1(\theta_i)$ favors alignment of 
the three successive sites $i-1$, $i$ and $i+1$. The nonbonded 
interactions $E_2$ are Lennard-Jones potentials with  
a species-dependent coefficient $C(\xi_i,\xi_j)$, which is       
taken to be 1 for an $AA$ pair (strong attraction), 
1/2 for a $BB$ pair (weak attraction) and -1/2 
for an $AB$ pair (repulsion). Consequently,  
there is an energetic preference for separation between 
the two kinds of monomers. In fact, it was demonstrated in
Ref.~\cite{Stillinger:93} that ground-state configurations 
tend to have a core consisting mainly of $A$ monomers, 
which shows that $A$ and $B$ monomers behave respectively as 
hydrophobic and polar residues.

The energy is, for fixed sequence, a function only of  
the angles $\theta_i$, and is therefore translationally and   
rotationally invariant. In addition, it is invariant under 
reflection ($\theta_i\to-\theta_i$ for all $i$) and
change of the orientation ($\theta_i\to-\theta_{N-i+1}$ 
and $\xi_i\to\xi_{N-i+1}$ for all $i$). This implies that
a generic energy level is fourfold degenerate for symmetric
sequences, and twofold degenerate for asymmetric sequences.   
In numerical simulations these symmetries provide 
useful checks on effective ergodicity breaking.        

The behavior of the model at finite temperature $T$ is defined 
by the partition function 
\beq
Z(T,\xi)=\int 
\biggl[\prod_{i=2}^{N-1}d\theta_i\biggr] 
\exp(-E(\theta,\xi)/T)\ .
\label{z}\eeq
That the $T$ dependence of this simple model exhibits an interesting 
phase structure will become clear in Secs. 4 and 5. 

\section{THE ALGORITHMS}

\subsection{Hybrid Monte Carlo}

We give a brief description of the 
hybrid Monte Carlo algorithm~\cite{Duane:87}, which is a general
method for simulating systems with continuous degrees of freedom
at constant temperature. More about the use of this method  
in simulating polymers and proteins can be found in 
Refs.~\cite{Brass:93,Heermann:93,Irback:94,Forrest:94}. 

The algorithm is based on the evolution arising from 
a fictitious Hamiltonian, which in our calculations
was taken as
\beq  
\Hmc(\pi,\theta)={1\over 2}\sum_{i=2}^{N-1}\pi_i^2 + E(\theta,\xi)/T
\label{hmc}\eeq
where $\pi_i$ is an auxiliary momentum variable conjugate to  
$\theta_i$. The first step in the algorithm is to generate a 
new set of momenta $\pi_i$ from the equilibrium distribution 
$P(\pi_i)\propto \exp(-{1\over 2}\pi_i^2)$. Starting from these 
momenta and the old configuration, the system is evolved through 
a finite-step approximation of the equations of motion. 
The configuration generated in such a trajectory is finally 
subject to an accept-or-reject question. The probability of 
acceptance in this global Metropolis step is
$\min(1,\exp(-\Delta\Hmc))$, 
where $\Delta\Hmc$ is the energy change in the trajectory. 
This accept-or-reject step removes errors due to the  
discretization of the equations of motion.

When integrating the discretized equations of motion, two sites
of the chain can come so close to each other that the 
hard $r^{-12}$ repulsion causes numerical instabilities. 
To avoid this problem, we have used a modified, linear 
potential for very small monomer-monomer separations. 
The effect of this on the sampling distribution was negligible, 
as the original and modified Boltzmann weights were both extremely 
small in this part of configuration space. In our 
simulations none of the accepted configurations belonged 
to the region with modified potential.     

The algorithm described above has two tunable parameters, namely the step 
size, $\epsilon$, and the number of steps in each trajectory, $n$. 
In our calculations we have used trajectories of length $n\epsilon=1$
and values of $n$ between 50 and 125.  

It has been shown how hybrid Monte Carlo can be used to simulate
efficiently various homopolymers with self-repulsion \cite{Irback:94}. 
However, in the application considered here, we expect the  
efficiency of hybrid Monte Carlo to be similar to that of 
conventional methods. 
 
\subsection{The dynamical-parameter method}

A simulation method meant for systems with a rough
free-energy landscape is the dynamical-parameter method. The basic step in 
this approach is to create an enlarged configuration space by introducing 
a new variable $k$. This variable can take $K$ different values, 
$k=1,\ldots,K$, which correspond to different values of the parameters 
of the model. In this way one tries to circumvent the barriers 
in the original free-energy landscape. 

We use the dynamical-parameter method to simulate the Boltzmann 
distribution $P_{\xi,T}(\theta)\propto \exp(-E(\theta,\xi)/T)$
for fixed sequence $\xi$ and temperature $T$. 
To do that, we assign one sequence, $\xi^{(k)}$, and one temperature, 
$T^{(k)}$, to each $k$, taking $\xi^{(k)}=\xi$ 
and $T^{(k)}=T$ for some $k$. Using standard methods, we then simulate 
the joint probability distribution $P(\theta,k)\propto 
\exp(-g_k-E(\theta,\xi^{(k)})/T^{(k)})$, 
where the $g_k$'s are constants that will be discussed below. 
Finally, knowing $P(\theta,k)$, one can obtain the  
distribution $P_{\xi^{(k_0)},T^{(k_0)}}(\theta)$, for any $k_0$, 
by simply restricting $P(\theta,k)$ to the subspace of fixed $k=k_0$. 

We study two different algorithms of this type,
simulated tempering~\cite{Marinari:92} and the multisequence algorithm.
The difference between these algorithms lies in the 
choice of the sets $\{\xi^{(k)}\}$ and $\{T^{(k)}\}$.   
In simulated tempering the temperature
fluctuates, while the sequence is held fixed, i.e.,   
$\xi^{(1)}=\ldots=\xi^{(K)}$. The multisequence algorithm,
on the other hand, uses
a set of different sequences and a fixed temperature, 
i.e., $T^{(1)} = \ldots = T^{(K}{}^)$.

The $g_k$'s are free parameters that determine the weights $p_k$ of the 
different $k$ values, which are given by 
\beq
p_{k}={ \exp(-g_{k})Z_{k}\over
\sum_{k^\prime=1}^K\exp(-g_{k^\prime})Z_{k^\prime}}
\label{pk}\eeq
where $Z_k=Z(T^{(k)},\xi^{(k)})$.
The method is free from systematic errors for any 
choice of the $g_k$'s, but the efficiency depends  
strongly on these parameters. In fact, if they are not 
chosen carefully, it can easily happen that the system gets 
trapped at a fixed value of $k$. In our calculations  
the $g_k$'s have been chosen so as to obtain a roughly uniform 
$k$ distribution. To that end, we used a number of trial 
runs, as will be described in Sec.~6. If the $p_k$'s 
corresponding to some choice of the $g_k$'s are known, 
the uniform distribution is obtained by replacing 
$g_k$ by $g_k+\ln p_k$.
 
Next we turn to our simulations of the joint distribution   
$P(\theta,k)$. For the $\theta$ update we have employed 
the hybrid Monte Carlo method described above. To facilitate  
comparisons, we used the same values of $n$ and $\epsilon$ 
as in our hybrid Monte Carlo runs. Our updates of $k$ were 
ordinary Metropolis steps \cite{Metropolis:53}. In our simulated-tempering
runs we used an ordered set of allowed temperature values and
changed $k$ in steps of $\pm 1$. Each hybrid Monte Carlo trajectory
was followed by one Metropolis step in $k$, and the average 
acceptance rate was 79\% for the $k$ update. In our multisequence runs  
we chose to work with all possible sequences with a 
fixed composition, i.e., with fixed numbers of $A$ and $B$ 
monomers, and the updates of $k$ corresponded to  
interchanges of the monomers at two randomly chosen sites. 
Here one update cycle consisted of one hybrid Monte Carlo trajectory 
followed by 100 Metropolis steps in $k$. In the final run, after 
adjusting the $g_k$'s, about 4\% of the cycles led to a change of $k$.       

The acceptance rate for the $\theta$ update depends on $k$ and was 
typically around 90\%. The fraction of the total computer time spent on 
$k$ updates was about 12\% in our multisequence calculations  
and much less than 1\% in our simulated-tempering calculations. Notice 
that the $k$ update leave the energy unchanged in simulated tempering.

In simulated tempering it is possible to avoid a low acceptance 
rate for the $k$ update by choosing a sufficiently dense (and 
large) set of $T^{(k)}$'s. The situation is different 
for the multisequence algorithm, since the sequences $\xi^{(k)}$ 
are picked from a discrete set. If the acceptance rate is low
in a multisequence simulation, it may be useful to introduce 
auxiliary sequences that interpolate between those of the model studied.    

\section{HOMOPOLYMERS}
  
We begin our study of the model defined in Sec.~2 by investigating
the behavior of homopolymers and how that depends on the temperature. 
The numerical simulations discussed in this and the next section  
have been performed by using the hybrid Monte Carlo method.  

It is well-known that the importance of the $r^{-6}$ terms in 
the potential varies with the temperature. At high temperature 
the influence of these terms is weak and the chains are 
expected to behave essentially as self-avoiding walks. 
At low temperature, on the other hand, they give rise to a strong effective 
attraction between monomers of the same type, 
which leads to globular chains.   

\begin{figure}[tbp]
\begin{center}
\vspace{-42mm}
\mbox{\hspace{-31mm}\psfig{figure=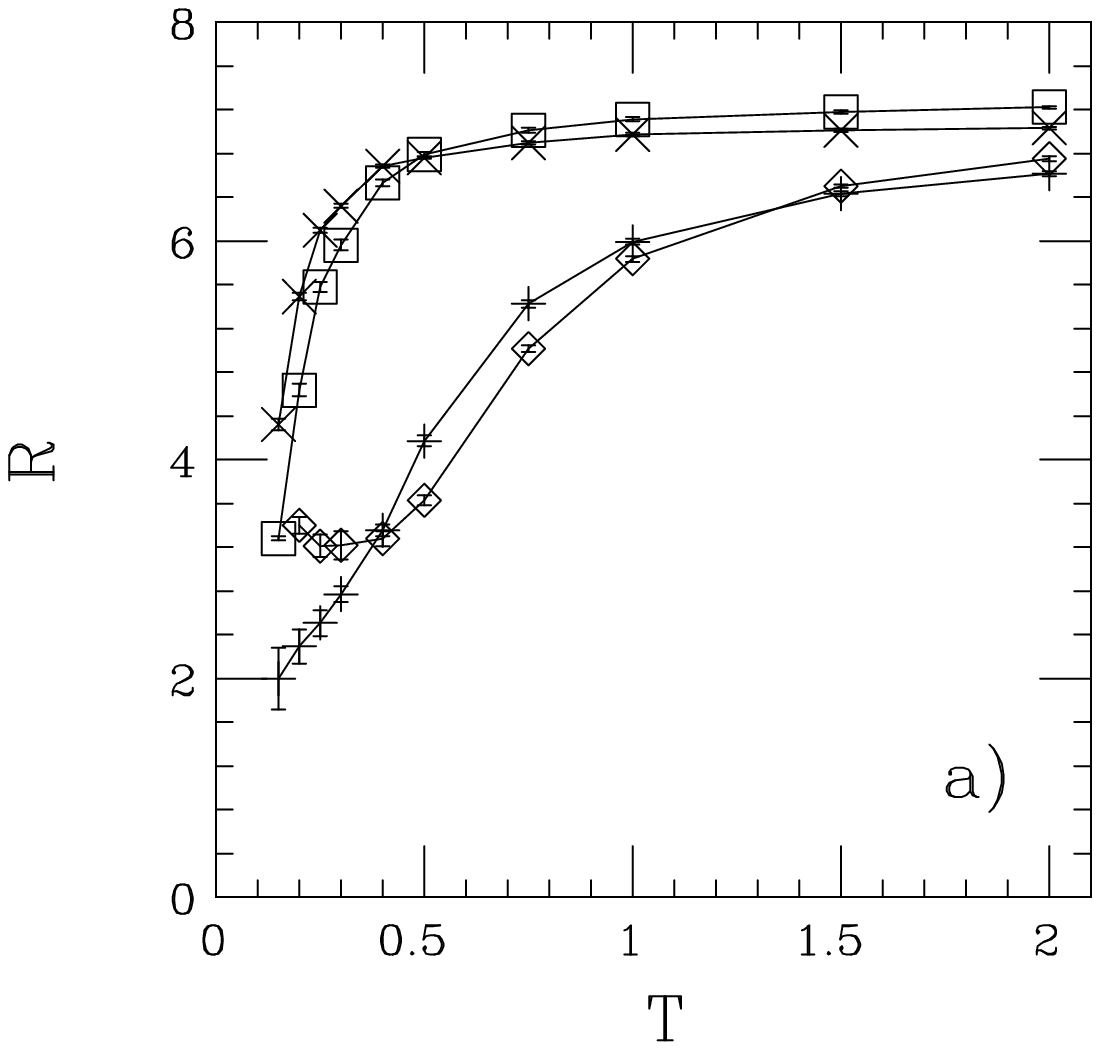,width=10.5cm,height=14cm}
\hspace{-30mm}\psfig{figure=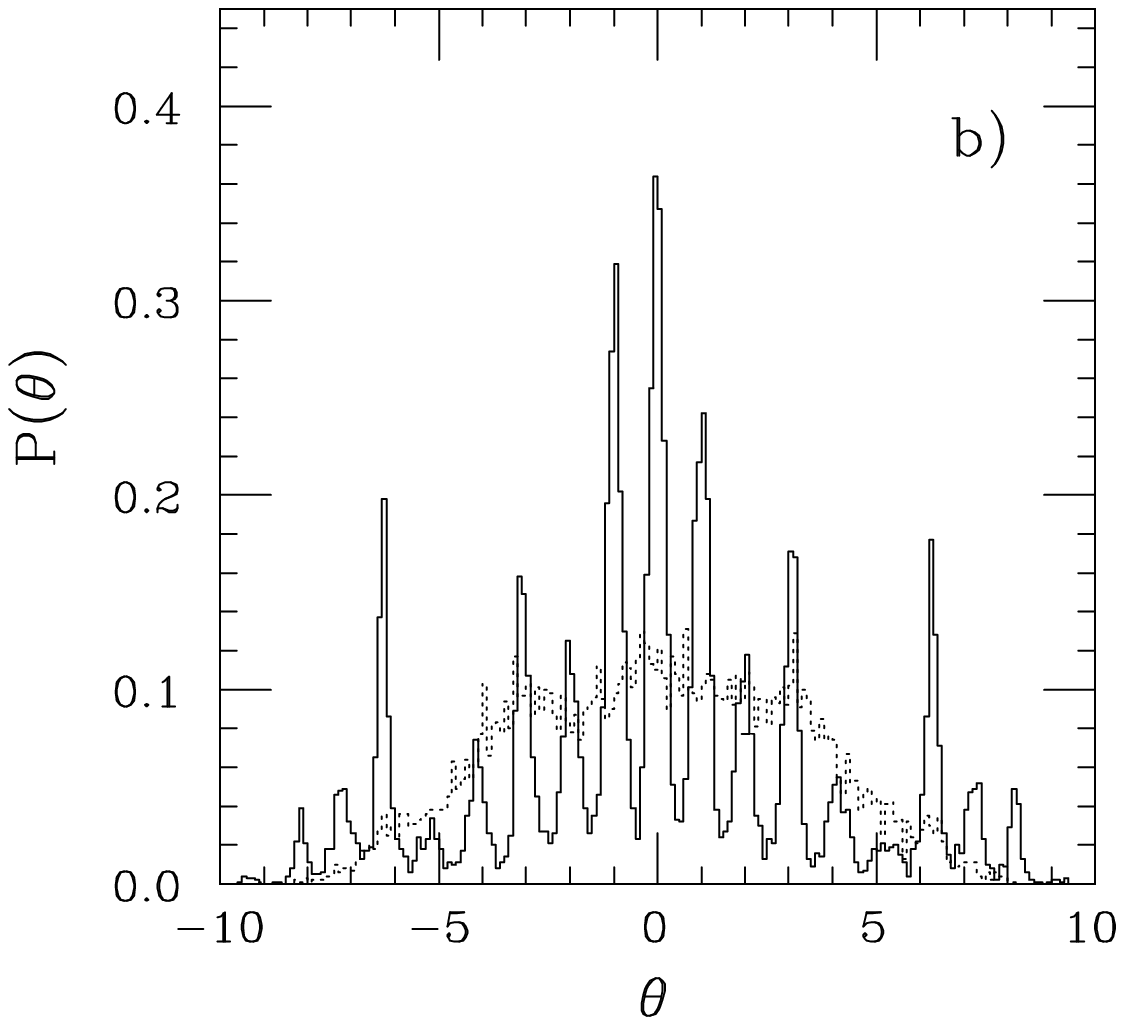,width=10.5cm,height=14cm}}
\vspace{-42mm}
\end{center}
\caption{a) The temperature dependence of the ratio 
$R=\ev{r_{ee}^2}/\ev{r_{gyr}^2}$ for $A$ (pluses $N=8$, diamonds 
$N=12$) and $B$ (crosses $N=8$, squares $N=12$) homopolymers.
b) Histogram of the total bend angle $\theta$ for 
$A$ (solid line) and $B$ (dotted line) homopolymers with $N=12$ at 
$T=0.2$.}    
\label{fig:1}
\end{figure}

To study how the extent of the homopolymers varies with the temperature,
we carried out numerical simulations for $N=8$ and 12. 
In Fig.~\ref{fig:1}a we show the results for the ratio 
$R=\ev{\ree^2}/\ev{\rgyr^2}$, where $\ree$ denotes the end-to-end 
distance and $\rgyr$ the radius of gyration. At high temperature 
the results are fairly close to $R\approx 7.13$, which is the 
value expected for two-dimensional self-avoiding walks in the limit 
$N\to\infty$ \cite{Cloizeaux:90,Li:94}. The corresponding value for 
ideal and quasi-ideal chains is $R=6$. Values of $R$ less than 6 
indicate that the system is in the globule phase. From the figure we estimate
that both the homopolymers are in the globule phase if $T<0.3$.
Our study of general sequences, which will be discussed in the next 
section, has been carried out using $T=0.2$. 

\begin{figure}[t]
\begin{center}
\vspace{-42mm}
\mbox{\hspace{-31mm}\psfig{figure=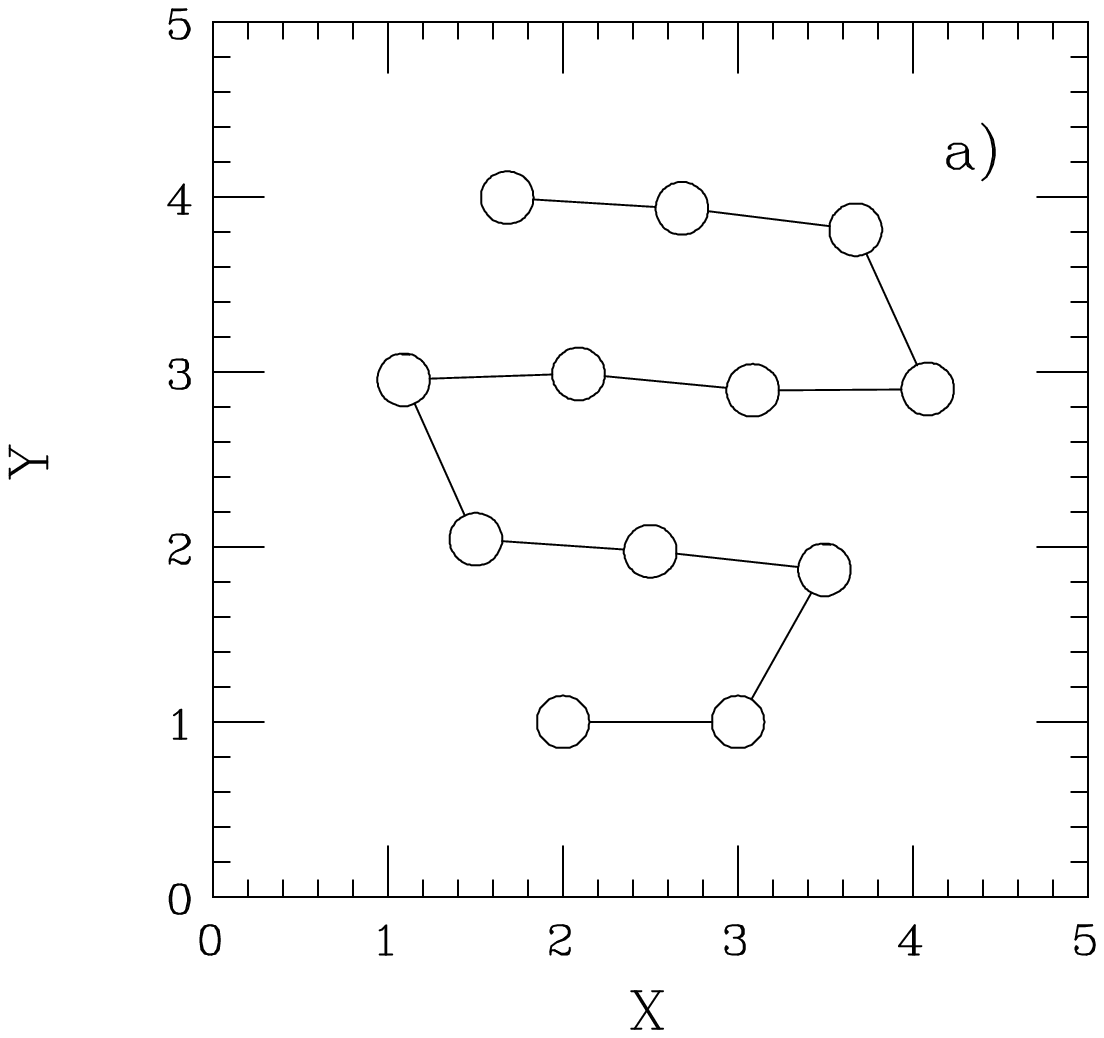,width=10.5cm,height=14cm}
\hspace{-30mm}\psfig{figure=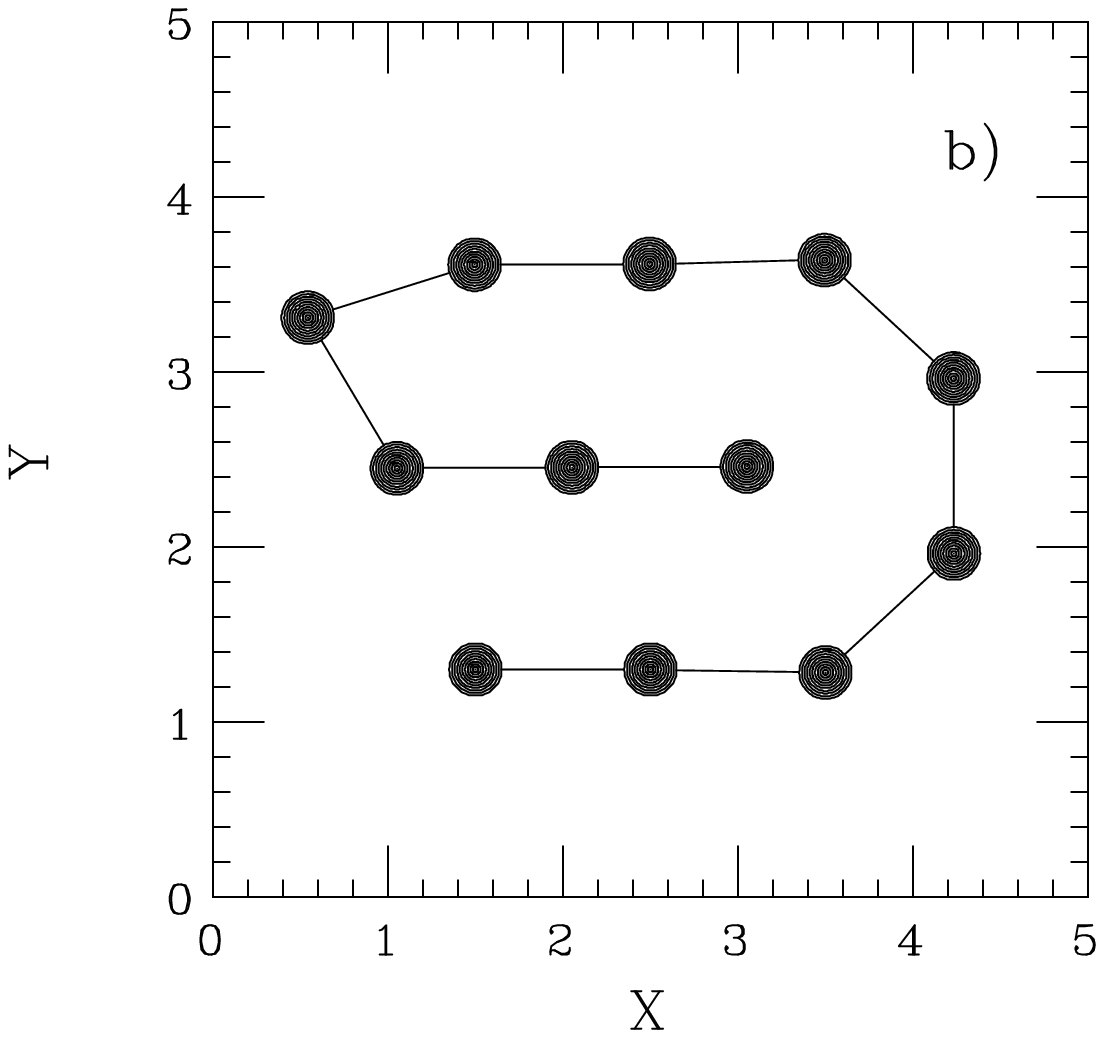,width=10.5cm,height=14cm}}
\vspace{-42mm}
\end{center}
\caption{Ground-state configurations for the a) $A$ and b) $B$   
homopolymers with $N=12$.}
\label{fig:2}
\end{figure}

Even though $A$ and $B$ homopolymers are both compact at $T=0.2$, 
they behave slightly differently. This can be 
seen from the probability distributions of the total bend angle 
$\theta=\sum_{i=2}^{N-1}\theta_i$ which are shown in 
Fig.~\ref{fig:1}b. The distribution for the $A$ chain has many 
narrow peaks, while that for the $B$ chain is dominated by 
a single, broad peak.
The origin of this difference becomes clear when studying 
the low-lying energy minima. To determine local energy minima 
we have employed a quenching procedure; during a Monte Carlo
simulation at fixed temperature, the system was quenched 
to zero temperature at regular intervals 
by using a conjugate gradient
method. In Fig.~\ref{fig:2} we show the minimum energy configurations 
for $N=12$. After checking the results by repeated 
use of simulated annealing, we feel confident that indeed these 
are ground-state configurations. Also, we checked that 
all the four symmetry related copies of these states were 
visited in the simulations. 

The structure of the ground state is the result of an interplay
between the bend energy and the Lennard-Jones terms. In Fig.~\ref{fig:2}
the sites of the $A$ chain approximately reside on a regular triangular
lattice, which is not true for the $B$ chain. This reflects the fact
that the bend energy plays a more important role for the $B$ than 
for the $A$ chain, which has deeper Lennard-Jones potentials.
The lattice structure can also be seen 
in other low-lying states of the $A$ homopolymer, and is responsible
for the many distinct peaks in the $\theta$ distribution 
(see Fig.~\ref{fig:1}b).

\section{GENERAL SEQUENCES}

Our study of general sequences has been carried out using $N=8$ and 10 
and a fixed temperature $T=0.2$. At this temperature the homopolymers 
have a compact form, but large fluctuations take place in the 
positions of the individual monomers. As will be shown in this section,
there are mixed sequences which, by contrast, exist in a unique state 
of fairly well-defined shape.     
  
To get a measure of the fluctuations in shape, we introduce the 
usual mean-square distance between configurations. For two configurations
$a$ and $b$ we define
\beq
\delta^2_{ab}\lika\min {1\over N}\sum_{i=1}^N 
|\bar x^{(a)}_i-\bar x^{(b)}_i|^2
\label{delta}\eeq
where $|\bar x^{(a)}_i-\bar x^{(b)}_i|$ denotes the distance 
between the sites $\bar x^{(a)}_i$ and $\bar x^{(b)}_i$
($\bar x^{(a)}_i,\bar x^{(b)}_i\in R^2$), 
and where the minimum is taken over translations, rotations and 
the discrete symmetries discussed in Sec.~2.
The probability distribution of $\delta^2$ for fixed temperature
and sequence, $P(\delta^2)$, can be obtained
numerically and is very informative \cite{Iori:91}, as it describes
the magnitude of the thermodynamically relevant fluctuations.
To determine $P(\delta^2)$, we computed $\delta^2$ for every possible
pair in a set of 2000 configurations, recorded at intervals of 
100 trajectories or more.  

\begin{figure}[tbp]
\begin{center}
\vspace{-42mm}
\mbox{\hspace{-31mm}\psfig{figure=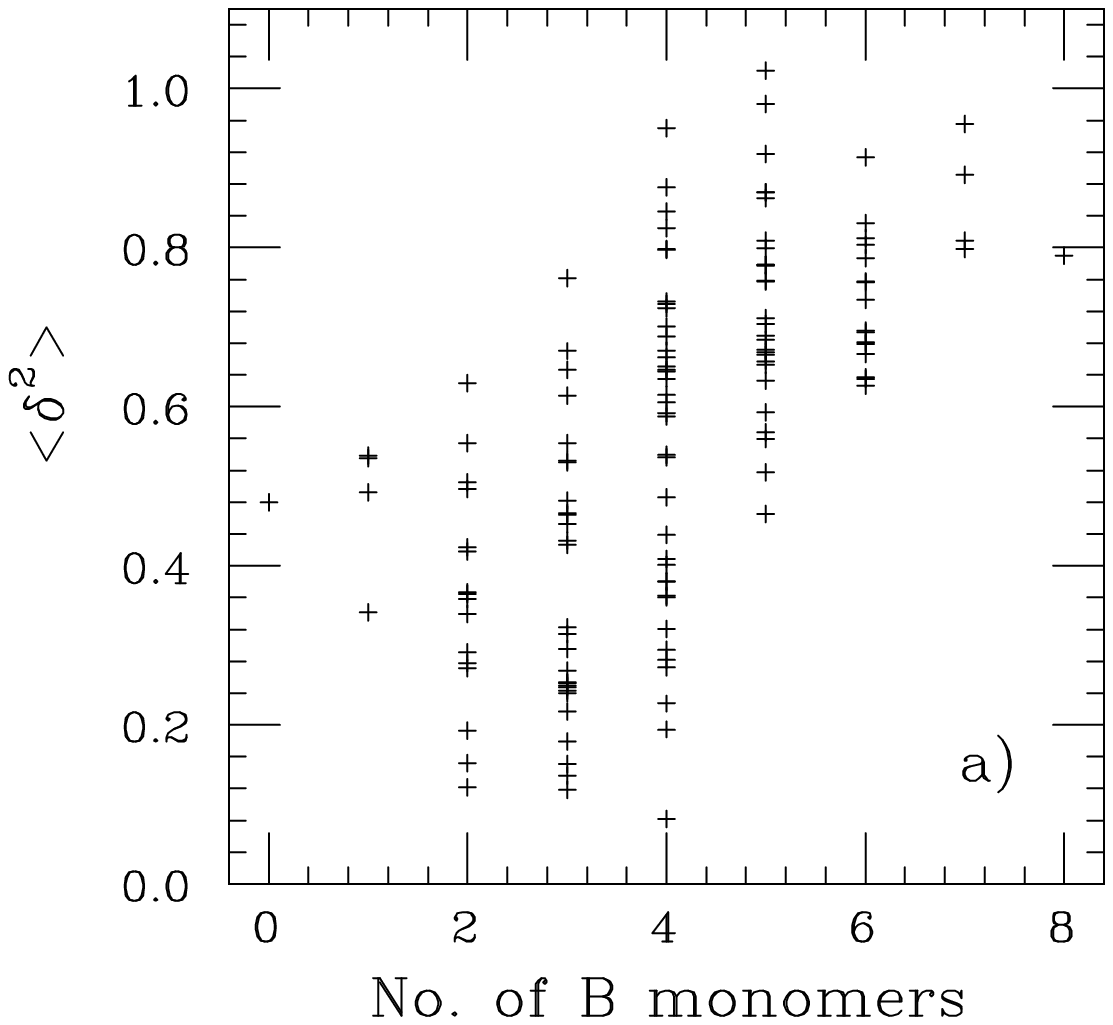,width=10.5cm,height=14cm}
\hspace{-30mm}\psfig{figure=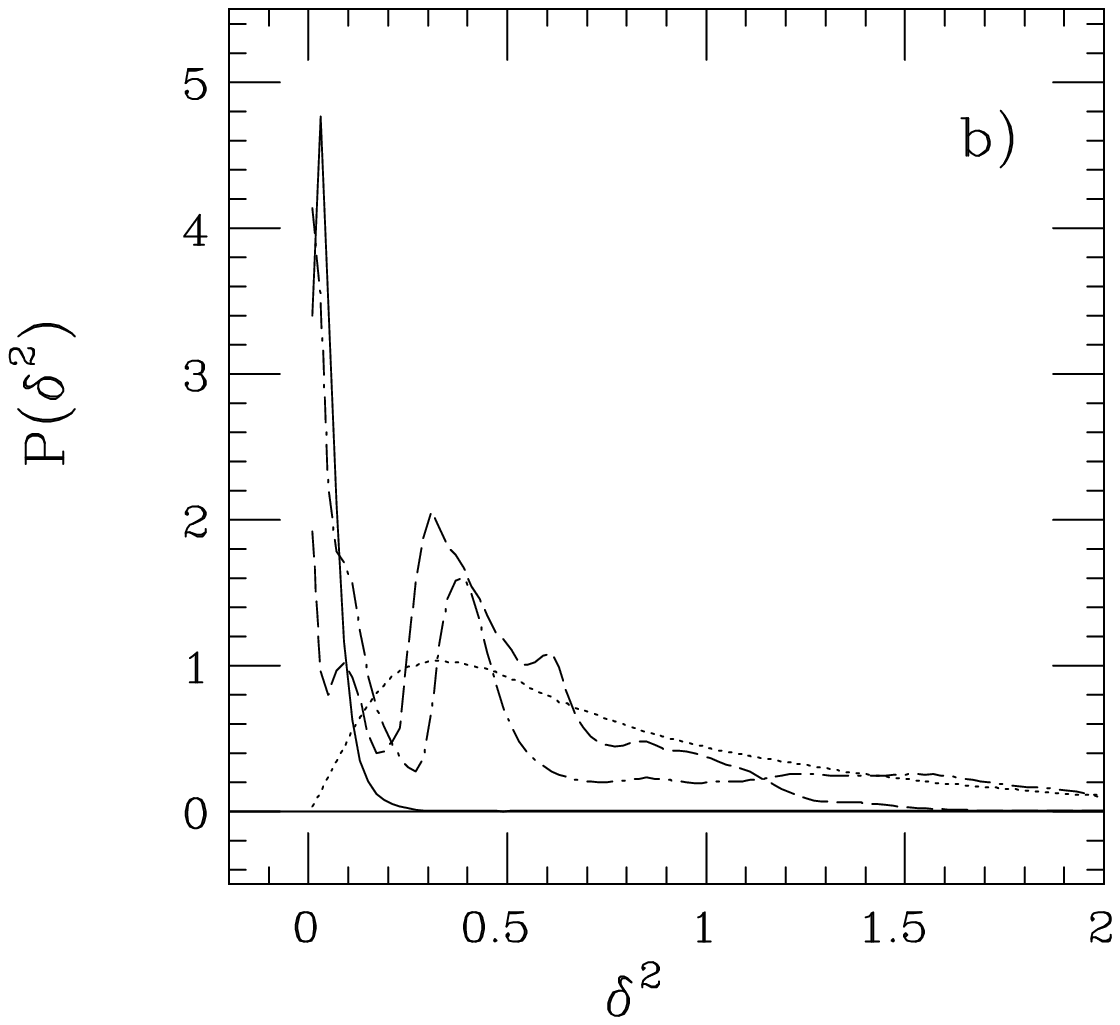,width=10.5cm,height=14cm}}
\vspace{-42mm}
\end{center}
\caption{
a) Average values $\ev{\delta^2}$ for the 136 chains with $N=8$,
plotted against the number of $B$ monomers. The statistical errors
are roughly of the size of the plot symbols.  
b) Histograms of $\delta^2$ for the sequences S1 (dashes), 
S2 (dots), S3 (solid) and S4 (dot-dash) (see Table 1).
The distribution for S3 has been normalized to 1/3.} 
\label{fig:3}
\end{figure}

We have calculated $P(\delta^2)$ for all the 136 distinguishable
chains with $N=8$ (the number of distinguishable
$2n$-mers is $2^{n-1}(2^n+1)$). 
In Fig.~\ref{fig:3}a we show the average values $\ev{\delta^2}$ 
of these distributions, plotted against the number of $B$ monomers. 
This figure shows that $\ev{\delta^2}$ varies considerably, and that values 
occur which are much smaller than those for the homopolymers. 
In Fig.~\ref{fig:3}b we show $P(\delta^2)$ 
for the sequences S1--4 in Table~\ref{tab:1}. 
S1 and S2 are the two homopolymers. 
S3 has the smallest average value
$\ev{\delta^2}$ of all the 136 sequences. S4 has the same composition 
as S3 but a more typical $\ev{\delta^2}$. 
That the shape of $P(\delta^2)$ is irregular for the $A$ homopolymer    
is due to the lattice structure discussed above. 
It should become smoother with increasing $N$.  
For S3 the distribution is dominated by a single, narrow peak located at 
low $\delta^2$, which shows that this chain exists in a 
state of fairly precise shape. There are other $N=8$  
chains which exhibit the same type of behavior.
However, the number of such sequences is small, as can be seen 
from the average values in Fig.~\ref{fig:3}a. 

The folding temperature $T_f$ may be defined as the lowest temperature
at which $\ev{\delta^2}$ takes some suitably chosen value $c$.
For reasonable choices of $c$, it is clear from Fig.~\ref{fig:3}a 
that the number of $N=8$ sequences with $T_f>0.2$       
is small.

\begin{table}[b]
\begin{center}
\begin{tabular}{|l|l|l|l|l|} \hline
   & $N$ & sequence     & $\ev{\delta^2}$ & $\ev{\rgyr^2}$ \\ \hline
S1 &  8  & AAAAAAAA     & 0.48(2)  & 1.262(3) \\ 
S2 &     & BBBBBBBB     & 0.79(2)  & 2.31(2)  \\ 
S3 &     & AABBBBAA     & 0.081(6) & 1.555(5) \\ 
S4 &     & ABAABBBA     & 0.57(2)  & 1.919(2) \\ \hline
S5 & 10  & AAAAAAAAAA   & 0.778(6) & 1.633(2) \\ 
S6 &     & BBBBBBBBBB   & 1.164(12)& 2.941(11)\\ 
S7 &     & AABAABAABA   & 0.090(8) & 1.707(2) \\ 
S8 &     & ABAAAABABA   & 0.510(12)& 1.827(2) \\ 
\hline 
\end{tabular}
\caption{Sequences referred to in the text. The average values 
were obtained at $T=0.2$.}
\label{tab:1}
\end{center}
\end{table}

\begin{figure}[tbp]
\begin{center}
\vspace{-42mm}
\mbox{\hspace{-31mm}\psfig{figure=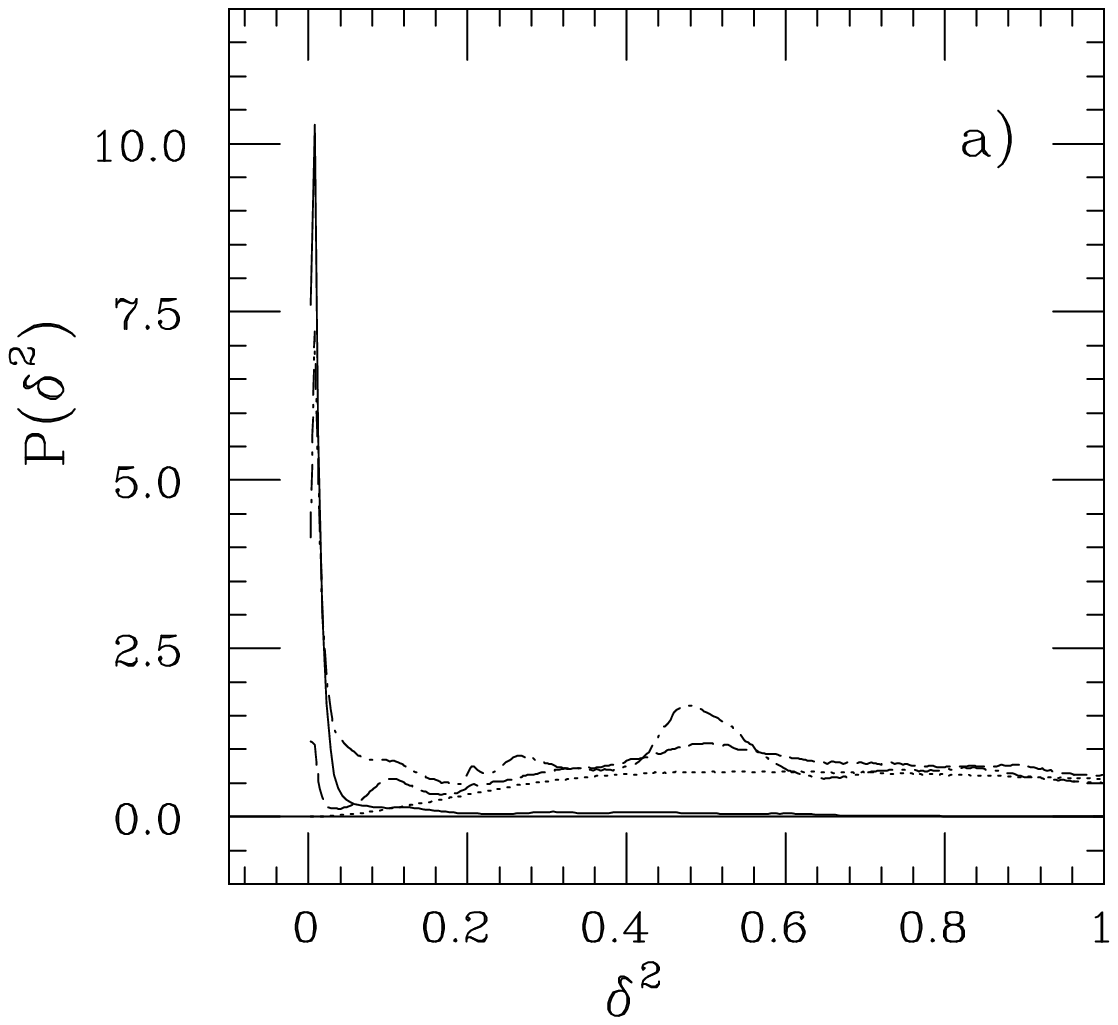,width=10.5cm,height=14cm}
\hspace{-30mm}\psfig{figure=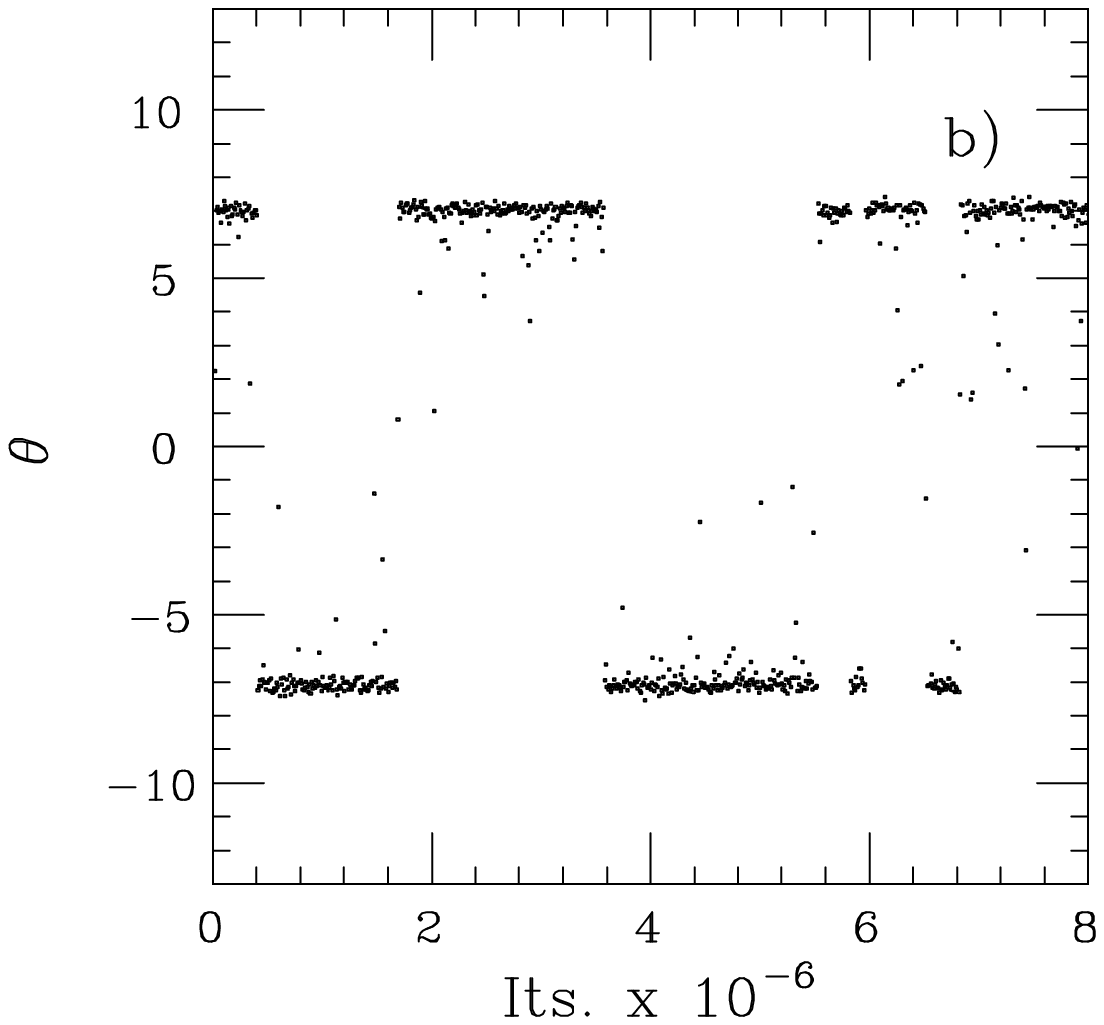,width=10.5cm,height=14cm}}
\vspace{-42mm}
\end{center}
\caption{a) Histogram of $\delta^2$ for the sequences 
S5 (dashes), S6 (dots), S7 (solid) and S8 (dot-dash) (see Table 1). 
The distribution for S7 has been normalized to 1/5. 
b) Monte Carlo evolution of the total bend angle  
$\theta$ for the chain S7. Data have been   
taken every 10,000 trajectories.}     
\label{fig:4}
\end{figure}

To test the size dependence of these results, we have also
studied the chains containing ten monomers. First we performed 
short simulations of all the 528 chains of this length. 
Even though some of these simulations were too short, the results 
clearly showed that most of the chains do not   
exhibit a well-defined shape at this temperature. 
Notice that too short runs tend to give underestimates of 
$\ev{\delta^2}$. After these preliminary runs, 
we picked out the four sequences S5--8 in Table~\ref{tab:1} 
and performed longer simulations of these. S7 is the $N=10$ sequence
with smallest $\ev{\delta^2}$, according to the preliminary runs,
and S8 is a sequence with the same composition as S7.   
In Fig.~\ref{fig:4}a we show $P(\delta^2)$ for S5--8. The result
of the longer run confirms that S7 exists in a unique folded
state. Since the sequence is asymmetric, there are two copies
of this state. From Fig.~\ref{fig:4}b it can be seen that
these symmetry related states were both visited in the simulation,
but transitions from one state to the other are rare.  

\begin{figure}[t]
\begin{center}
\vspace{-42mm}
\mbox{\hspace{-31mm}\psfig{figure=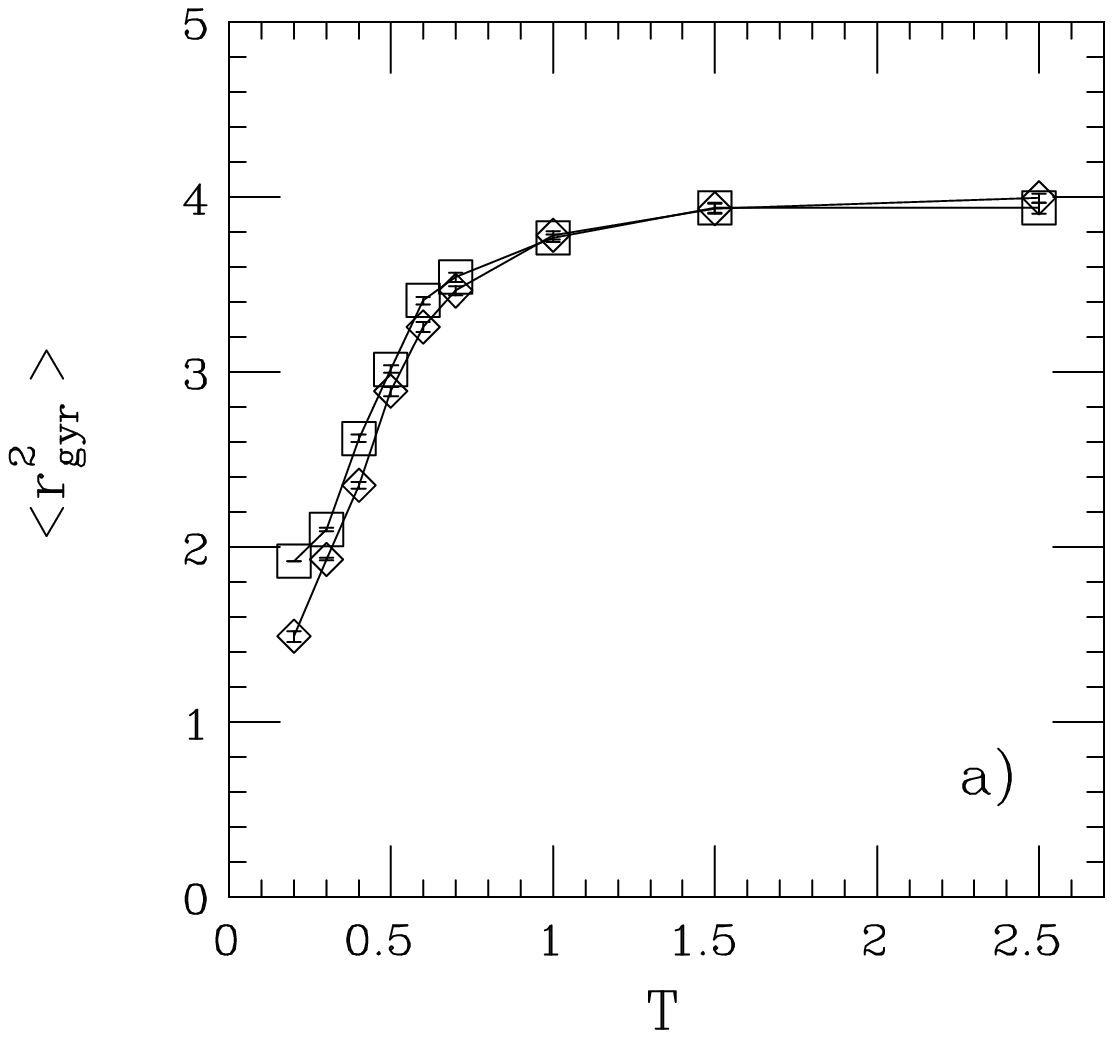,width=10.5cm,height=14cm}
\hspace{-30mm}\psfig{figure=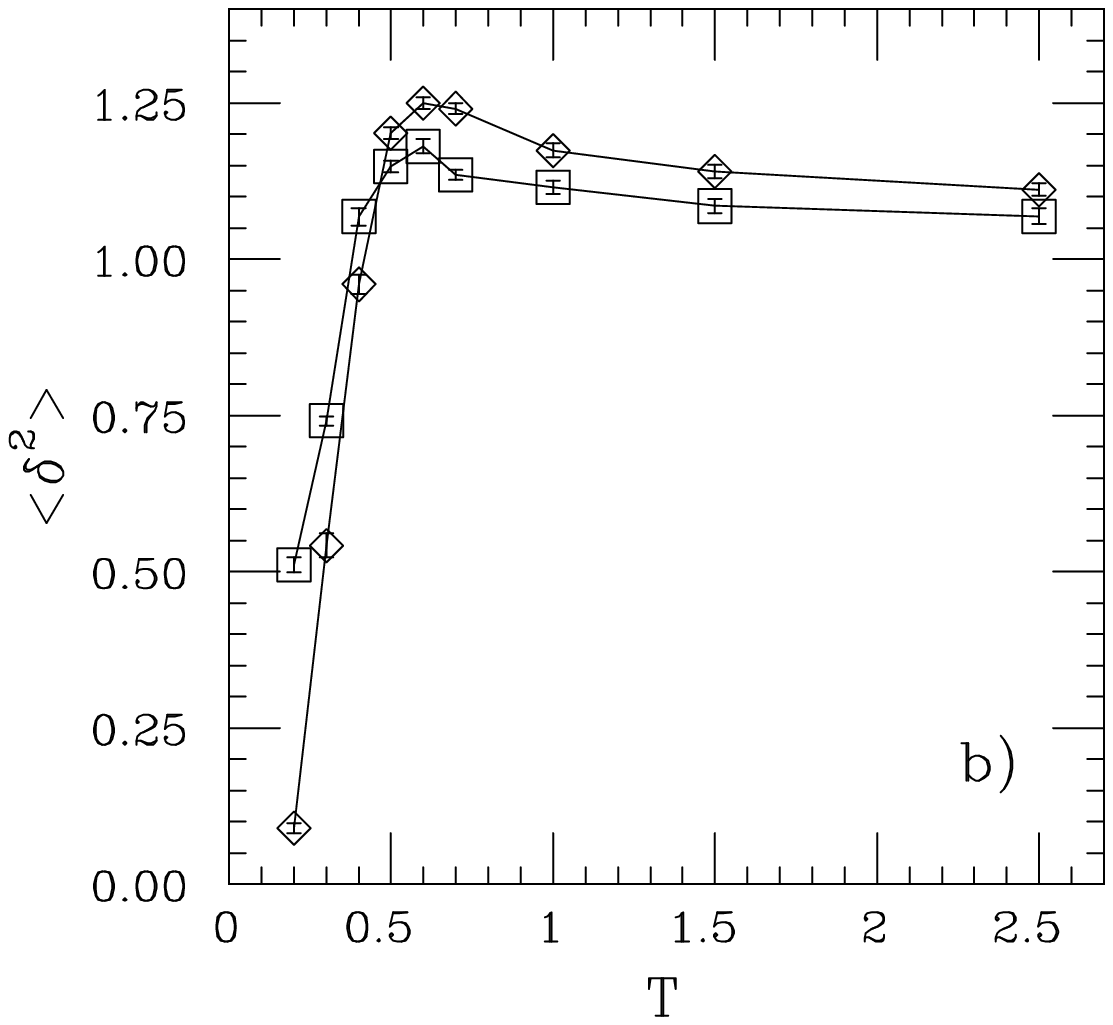,width=10.5cm,height=14cm}}
\vspace{-42mm}
\end{center}
\caption{Temperature dependence of
a) $\ev{r_{gyr}^2}$ and b) $\ev{\delta^2}$ for the sequences
S7 (diamonds) and S8 (squares) (see Table~1).}
\label{fig:5}
\end{figure}

The results presented so far in this section were obtained 
at a fixed temperature. Let us also briefly discuss the transition  
to the folded low-temperature phase for the sequence S7. We compare 
the behavior of S7 with that of S8, which has a folding
temperature lower than 0.2. In Fig.~\ref{fig:5}a we show 
the temperature dependence of $\ev{\rgyr^2}$ 
and $\ev{\delta^2}$. The radius of gyration is similar for the
two sequences, which is expected since they have the same composition, 
and decreases gradually as the temperature is decreased.
The temperature dependence 
of $\ev{\delta^2}$ is more dramatic. We see that the transition from a 
high-temperature state with large fluctuations in shape to a 
frozen state of well-defined shape is fairly abrupt for S7.

\begin{figure}[t]
\begin{center}
\vspace{-91mm}
\mbox{\hspace{-18mm}
\psfig{figure=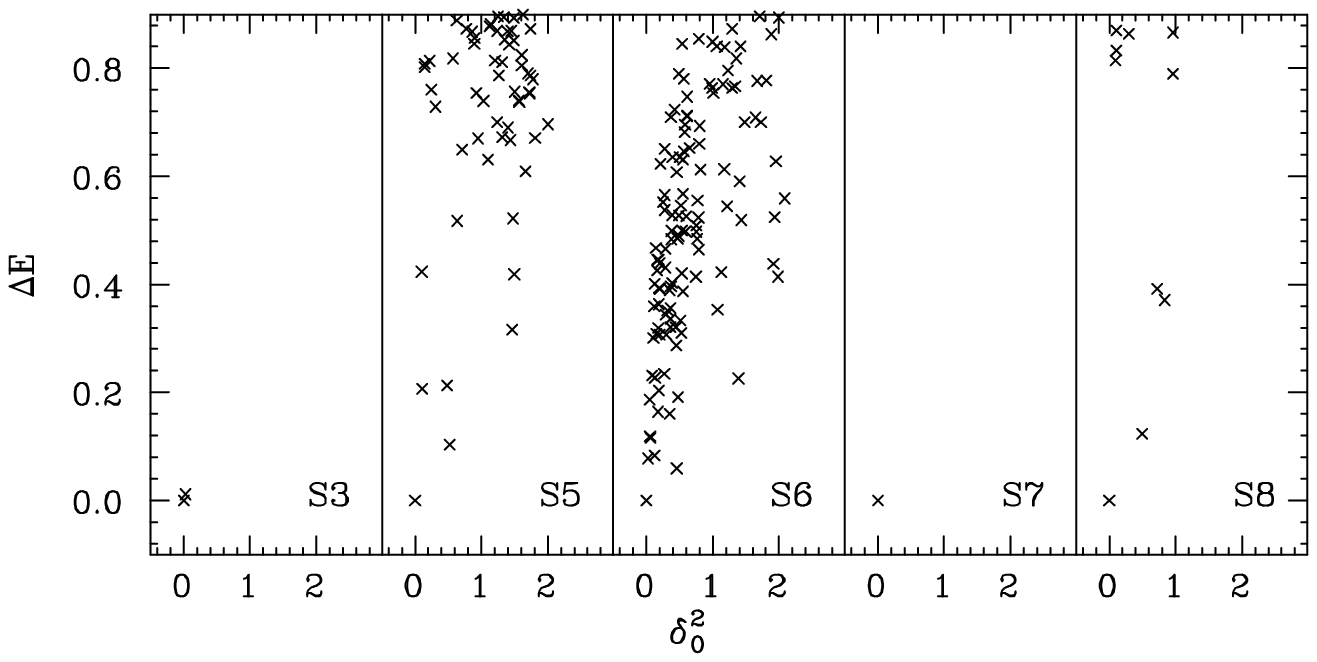,width=18cm,height=24cm}}
\vspace{-96mm}
\end{center}
\caption{Local energy minima for the sequences S3 and 
S5--8 (see Table~1).}
\label{fig:6}
\end{figure}

Finally, let us compare the energy level spectra of the sequences 
S3 and S5--8. We\break
have seen that S3 and S7, unlike the other three of these sequences, 
have a folding temperature higher than 0.2. In Fig.~\ref{fig:6} we 
show how the low-lying local energy minima are distributed in the 
$(\delta_0^2,\Delta E)$ plane, where $\delta_0^2$ is the mean-square 
distance to the lowest minimum found (cf Eq.~\ref{delta}) and $\Delta E$ is 
the energy difference to this minimum. The minima were determined by 
using the quenching procedure mentioned in Sec.~4, and the figure shows 
all the minima found with $\Delta E<0.9$. From the figure it can be 
seen that for S7 there is only one minimum with $\Delta E<0.9$. 
For S3 there are two such minima which are structurally very similar. 
The stability gap defined in Sec.~1 is therefore large for both 
the sequences with high $T_f$, as expected. The behavior of S3 illustrates 
that it is important to distinguish between this gap and the energy gap 
between the two lowest of all minima.    

\section{TESTS OF THE DYNAMICAL-PARAMETER ALGORITHMS}
 
We have tested the performance of the two dynamical-parameter 
algorithms described in Sec.~3 using several different
sequences. The results obtained for the different sequences were 
qualitatively very similar. For clarity we focus here on the 
results for a specific sequence, namely $AABAAABBAA$, at $T=0.17$.   
   
\begin{figure}[tbp]
\begin{center}
\vspace{-50mm}
\mbox{\hspace{20mm}\psfig{figure=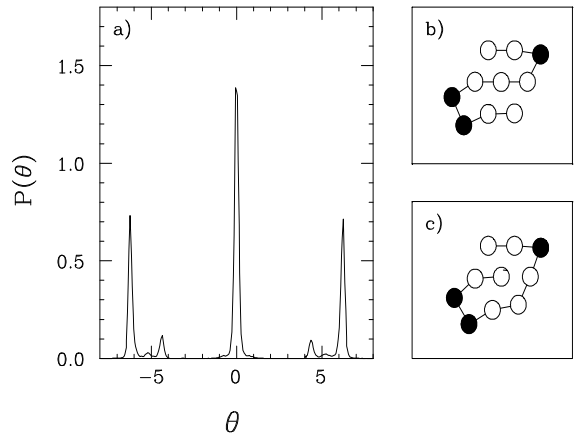,width=11cm,height=15.55cm}}
\vspace{-55mm}
\end{center}
\caption{a)~Histogram of the total bend angle $\theta$. 
b)~Ground-state configuration. c)~The next-lowest local 
energy minimum. Open and filled circles represent $A$ and $B$ monomers,
respectively.}    
\label{fig:7}
\end{figure}

In Fig.~\ref{fig:7}a we show the probability distribution of
the total bend angle $\theta=\sum_{i=2}^{N-1}\theta_i$, 
which is dominated by three narrow and well separated peaks. 
By measuring mean-square deviations (cf Eq.~\ref{delta}), we checked that   
the central peak represents small fluctuations around the
minimum energy configuration, which is shown in Fig.~\ref{fig:7}b. 
The next-lowest energy minimum is shown in Fig.~\ref{fig:7}c, 
and is very similar to the lowest minimum if one neglects
the chain structure. However, for the next-lowest minimum we have
$\theta\approx 6.1$, which coincides with the position of the 
rightmost peak in $P(\theta)$. The third peak, at $\theta\approx-6.1$,  
is related by symmetry to the one at $\theta\approx 6.1$.  

When simulating this system by standard methods, the major difficulty   
is that transitions between the three different regions corresponding to these 
peaks are strongly suppressed. By the dynamical-parameter method one 
can greatly improve the frequency of these transitions, as
will be seen below. Frequent transitions between the two symmetry related 
regions can, of course, also be obtained by incorporating global, random  
flips of the signs of the $\theta_i$'s into the update scheme. The important 
and nontrivial problem is to obtain transitions that connect these 
two regions with the $\theta\approx 0$ region.
   
\begin{figure}[tbp]
\begin{center}
\vspace{-50mm}
\mbox{\hspace{0mm}\psfig{figure=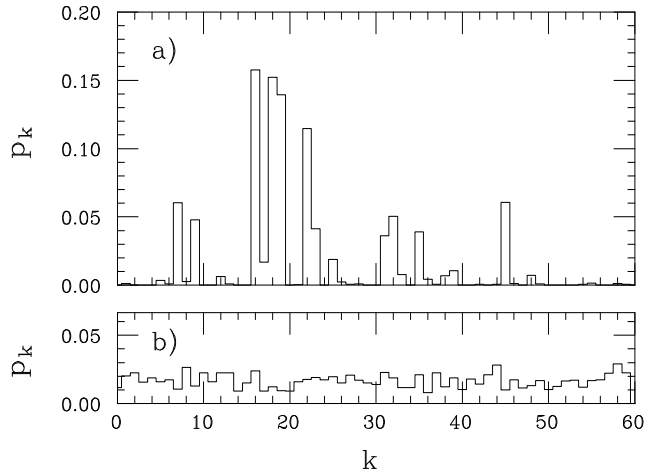,width=11cm,height=15.5cm}}
\vspace{-51mm}
\end{center}
\caption{The weights $p_k$ in the multisequence calculation: 
a) for all $g_k=0$ and b) for our final choice the $g_k$'s.}
\label{fig:8}
\end{figure}

Before running the dynamical-parameter algorithms, one has   
to determine the parameters $g_k$. We have done that in
slightly different ways for the two algorithms.  
In our multisequence runs the set of sequences $\xi^{(k)}$  
consisted of all the $60$ sequences  
containing seven $A$ and three $B$ monomers. 
To determine the corresponding $g_k$'s, we carried out 
four trial runs, covering a total of $6\times10^5$ update cycles. 
After each of these runs the $g_k$'s were adjusted, using the 
measured weights $p_k$ as described in Sec.~3. 
In Fig.~\ref{fig:8} we show the 60 different weights before and 
after the tuning procedure. None of the final $p_k$'s is  
smaller than $1/2\times 1/60\approx 0.008$. 

In our simulated-tempering runs there were 12 allowed values of the 
temperature, $0.17=T^{(1)}<\ldots<T^{(12)}=0.47$, which were  
equidistant in $1/T$. Here we first performed short 
hybrid Monte Carlo simulations, each consisting of 1000 trajectories, 
to get rough estimates of the average energy at these temperature values, 
$\ev{E}_k$. We then put $g_{12}=0$ 
and $g_{k-1}=g_k-\ev{E}_k(1/T^{(k-1)}-1/T^{(k)})$
for $k<12$. This relation between $g_{k-1}$ and $g_k$ is 
obtained by requiring that $p_{k-1}=p_k$ and neglecting
terms that are of order two or higher in $1/T^{(k-1)}-1/T^{(k)}$.
One of the $g_k$'s can be chosen freely, since only the
differences between them are relevant. To fine-tune the $g_k$'s,
we finally carried out two simulated-tempering runs, each 
consisting of $2\times10^4$ update cycles.

\begin{figure}[tbp]
\begin{center}
\vspace{-100mm}
\mbox{\hspace{-20mm}\psfig{figure=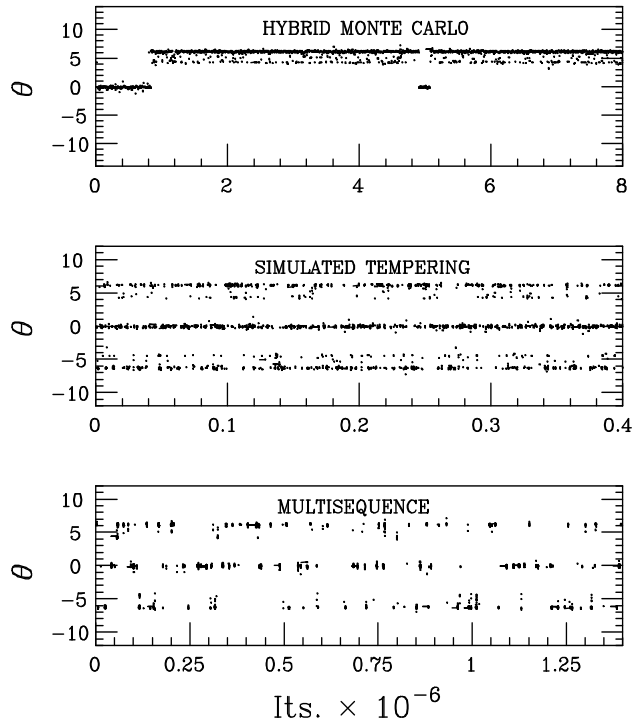,width=20cm,height=28.3cm}}
\vspace{-105mm}
\end{center}
\caption{Evolution of the total bend angle $\theta$ in three  
different simulations.}
\label{fig:9}
\end{figure}

Having chosen the $g_k$'s, we performed one long production run 
for each of the two algorithms. In Fig.~\ref{fig:9} we show the 
evolution of $\theta$ in these simulations. Only data corresponding 
to $T=0.17$ and sequence $AABAAABBAA$ are shown, and they are
plotted against the total number of update cycles. 
Also shown in Fig.~\ref{fig:9} is the result of a 
hybrid Monte Carlo run. The same values of the simulation 
parameters $n$ and $\epsilon$ have been used in the three 
different runs. The CPU time per iteration is therefore 
very similar for the hybrid Monte Carlo and simulated-tempering 
runs, and 12\% higher for the multisequence run. 

From Fig.~\ref{fig:9} it is evident that transitions between the
three different regions mentioned above are indeed much more 
frequent in the dynamical-parameter simulations than in the 
hybrid Monte Carlo simulation. In the hybrid Monte Carlo run
there are only six transitions 
(three of which are difficult to see in the figure) 
in $8\times 10^6$ trajectories,  
and one of the three regions is, in fact, never
visited. By contrast, every region is visited many
times in the two other runs, even though these are 
shorter. The frequency of transitions to and from the 
region with $\theta\approx 0$ is respectively $9.4\times 10^{-4}$ and
$5.7\times 10^{-5}$ per update cycle in the simulated-tempering
and multisequence runs. 
   
The improved transition frequency is the major advantage 
of the dynamical-parameter simulations. In addition, these simulations
can be used to study several different temperature values or sequences.
The disadvantage of the method is that  
a number of trial runs are needed in order to determine 
the $g_k$'s. In our simulated-tempering and multisequence 
calculations we have spent respectively 12\% and 30\% of 
the total computer time on trial runs. Let us stress that 
the total cost of each of these two calculations is still 
considerably lower than that of our hybrid Monte 
Carlo run (cf Fig.~\ref{fig:9}).
     
\section{Summary}

We have studied numerically the finite-temperature behavior of 
a simple off-lattice model for protein folding. All possible chains  
containing eight or ten monomers were studied at a fixed 
temperature, which was chosen low enough for all the chains to 
be fairly compact. We showed that most of the chains undergo large 
fluctuations in shape at this temperature, while a few 
exist in a unique state of well-defined shape. We determined the
low-lying local energy minima for a few different sequences, 
and found that the stability gaps vary considerably. This suggests 
that there are important differences in folding temperature, and, 
therefore, in the ability to fold fast, between different sequences.   

Although the relatively short chains studied here exhibit interesting 
properties, it would, of course, be desirable to extend the study 
to considerably longer chains. To be able to do that, it is of great 
interest to find improved Monte Carlo methods. We have in this paper 
studied two dynamical-parameter algorithms, simulated tempering and 
the multisequence algorithm, which were found to be much more efficient 
than conventional simulation methods. The two algorithms are also 
useful for energy minimization, if combined with a suitable 
local optimization method.  

It would be interesting to compare the performance of these 
algorithms with that of the multicanonical Monte Carlo algorithm which was 
mentioned in the introduction. Multicanonical heteropolymer simulations have
recently been reported~\cite{Hansmann:93,Hansmann:94,Hao:94}.

\newpage

\newcommand	{\Biopol}   {Biopolymers\ }
\newcommand	{\BC}       {Biophys.\ Chem.\ }
\newcommand	{\BJ}       {Biophys.\ J.\ }
\newcommand     {\COSB}     {Curr.\ Opin.\ Struct.\ Biol.\ }
\newcommand	{\EL}	    {Europhys.\ Lett.\ }
\newcommand     {\JCC}      {J.\ Comp.\ Chem.\ }
\newcommand	{\JCoP}	    {J.\ Comp.\ Phys.\ }
\newcommand	{\JCP}	    {J.\ Chem.\ Phys.\ }
\newcommand	{\JMB}	    {J.\ Mol.\ Biol.\ }
\newcommand	{\JP}	    {J.\ Phys.\ }
\newcommand     {\JPC}      {J.\ Phys.\ Chem.\ }
\newcommand	{\JSP}	    {J.\ Stat.\ Phys.\ }
\newcommand     {\M}        {Macromolecules\ }
\newcommand     {\MC}       {Makromol.\ Chem.,\ Theory Simul.\ }
\newcommand     {\MP}       {Molec.\ Phys.\ }
\newcommand     {\Nat}      {Nature}
\newcommand     {\NP}       {Nucl.\ Phys.}
\newcommand     {\Pro}      {Proteins:\ Struct.\ Funct.\ Genet.\ }
\newcommand     {\ProSci}   {Protein\ Sci.\ }
\newcommand     {\Pa}       {Physica\ }
\newcommand	{\PL} 	    {Phys.\ Lett.\ }
\newcommand	{\PNAS}     {Proc.\ Natl.\ Acad.\ Sci.\ USA\ }
\newcommand	{\PR}	    {Phys.\ Rev.\ }
\newcommand	{\PRL}	    {Phys.\ Rev.\ Lett.\ } 
\newcommand	{\Sci}	    {Science\ }
\newcommand	{\ZP}	    {Z.\ Physik\ }

\end{document}